\documentclass[conference]{IEEEtran}

\usepackage{amsmath,amssymb} 
\usepackage{color}           
\usepackage{graphicx}
\usepackage{caption}
\usepackage{subcaption}      
\usepackage{epstopdf}        
\usepackage{url}
\usepackage[ruled,linesnumbered]{algorithm2e}
\usepackage{enumerate}
\IEEEoverridecommandlockouts

\title{Cyber-Physical Attacks on UAS Networks- Challenges and Open Research Problems
}

\author{
  \IEEEauthorblockN{Vahid Behzadan}
  \IEEEauthorblockA{
    Dept. of Computer Science and Engineering, University of Nevada, Reno, USA\\
    vbehzadan@unr.edu
  }
}
\begin{document}
\maketitle

\begin{abstract}
Assignment of critical missions to unmanned aerial vehicles (UAV) is bound to widen the grounds for adversarial intentions in the cyber domain, potentially ranging from disruption of command and control links to capture and use of airborne nodes for kinetic attacks. Ensuring the security of electronic and communications in multi-UAV systems is of paramount importance for their safe and reliable integration with military and civilian airspaces. Over the past decade, this active field of research has produced many notable studies and novel proposals for attacks and mitigation techniques in UAV networks. Yet, the generic modeling of such networks as typical MANETs and isolated systems has left various vulnerabilities out of the investigative focus of the research community. This paper aims to emphasize on some of the critical challenges in securing UAV networks against attacks targeting vulnerabilities specific to such systems and their cyber-physical aspects.
\end{abstract}

\section{Introduction}\label{intro}
The 21st century is scene to a rapid revolution in our civilization's approach to interactions. Advancement of communication technologies, combined with an unprecedentedly increasing trust and interest in autonomy, are pushing mankind through an evolutionary jump towards delegation of challenging tasks to non-human agents. From mars rovers to search and rescue robots, we have witnessed this trend of overcoming the limitations inherent to us, through replacement of personnel with cyber-physical systems capable of performing tasks that are risky, repetitive,  physically difficult or simply economically infeasible for human actors.

Unmanned Aerial Vehicles, or UAVs, are notable examples of this revolution. Since the early 2000s, military and intelligence theaters have seen an explosive growth in the deployment of tactical UAVs for surveillance, transport and combat operations. In the meantime, civilian use of UAVs has gained traction as the manufacturing and operations costs of small and mid-sized UAVs are undergoing a steady decline. The cheaper cost of such UAVs has also led to a growing interest in collaborative deployment of multiple UAVs to perform specific tasks, such as monitoring the conditions of farms and patrolling national borders. Yet, there are a multitude of challenges associated with this vision, solving which are crucial for safe and reliable employment of such systems in civilian and military scenarios. One such challenge is ensuring the security of systems that comprise UAVs, as their remote operational conditions leave the burden of command and control reliant on the onboard cyber-physical components. The body of literature on this issue has seen an accelerated growth in recent years \cite{kim2012cyber}, which is partly due to major cyber attacks on UAVs \cite{javaid2012cyber}. The overwhelming number of potential vulnerabilities in UAVs indicates the need for vigorous standards and frameworks for assurance of reliability and resilience to malicious manipulations in all aspects of UAVs, from the mechanical components to the information processing units and communications systems.

In multi-UAV operations, Inter-UAV links are necessary for exchange of situational and operational commands, which are the basis of essential functions such as formation control and task optimization. As for the architecture of these UAV networks, the current consensus in the research community is biased towards decentralized and ad hoc solutions, which allow dynamic deployment of Unmanned Aerial Systems (UAS) with minimal time and financial expenditure on pre-mission preparations. 
\begin{figure}
	\centering
	\includegraphics[width=\linewidth]{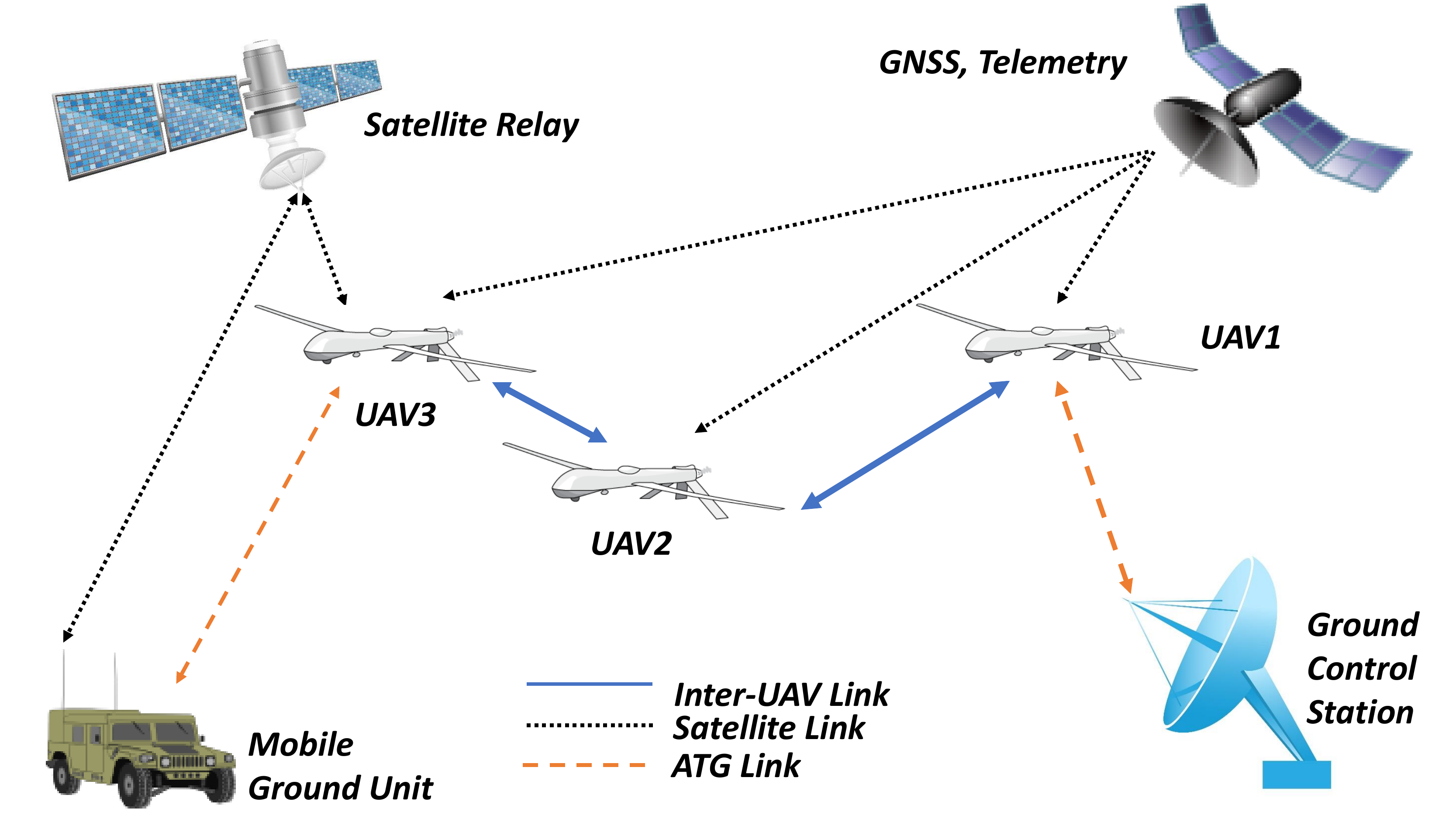}
	\caption{Communication Links in a UAS Network}
	\label{fig:FANET}
\end{figure}

Structure of a typical UAS network is shown in Figure~\ref{fig:FANET}. By considering the various types of links and interfaces depicted in this figure, it can be deduced that such networks are inherently of a complex nature. Integration of multiple subsystems not only aggregates their individual vulnerabilities, but may result in new ones that are rooted in the interactions between those subsystems. Hence, UAS present the research community with a novel interdisciplinary challenge. The aim of this paper is to emphasize on some of the critical vulnerabilities specific to network and communications aspects of UAVs, and provide the research community with a list of open problems in ensuring the safety and security of this growing technology.

\section{Uniqueness of UAS Networks}
Accurate analysis of vulnerabilities in UAS networks necessitates an understanding of how an airborne network differs from traditional computer networks. Much of recent studies in this area compare UAS networks to Mobile Ad hoc Networks (MANETs) and Wireless Sensor Networks (WSN), as UAS communications and protocols may initially seem similar to those of generic distributed and mobile networks. Yet differences in mobility and mechanical degrees of freedom, as well as their operational conditions, build the grounds for separate classification of UAS networks. One such distinguishing factor is the velocity of airborne vehicles, which may range up to several hundreds of miles per hour. The high mobility of airborne platforms increases the complexity of requirements for the communications subsystem and many aspects of the UAS network. In the link layer, management of links and adaptation of access control has to be fast enough to accommodate tasks such as neighbor discovery and resource allocation in an extremely dynamic environment. Likewise, the network layer must be able to provide fast route discovery and path calculation while preserving the reliability of the information flow.

In the physical layer, not only communications, but the kinetic aspects of the UAS give rise to unique requirements. As the span of a UAS network may vary from close-by clusters to far and sparse distributions, the transmission power of UAV radios must be adjustable for efficient power consumption and sustained communications. Also, since the geography and environment of the mission may vary rapidly, channel availability in UAS links is subject to change. A potential solution is for the UAS to be equipped with Dynamic Spectrum Access (DSA) and adaptive radios to provide the required agility. Furthermore, the conventional antenna arrangement on airborne platforms is such that changes in orientation and attitude of the aircraft affect the gain of onboard radios. This problem is further intensified in unmanned aircraft, as the elimination of risk to human pilot allows longer unconventional maneuvers.

These considerations clarify the demand for a fresh vantage point for analyzing the problem of security in UAS networks. The reliability of today's mission-critical UAVs need to be studied with models that adopt a more inclusive view of such systems and the impact of seemingly benign deficiencies on the overall vulnerability of UAVs.

\section{Anatomy of a UAV}

\begin{figure}
	\centering
	\includegraphics[width=\linewidth]{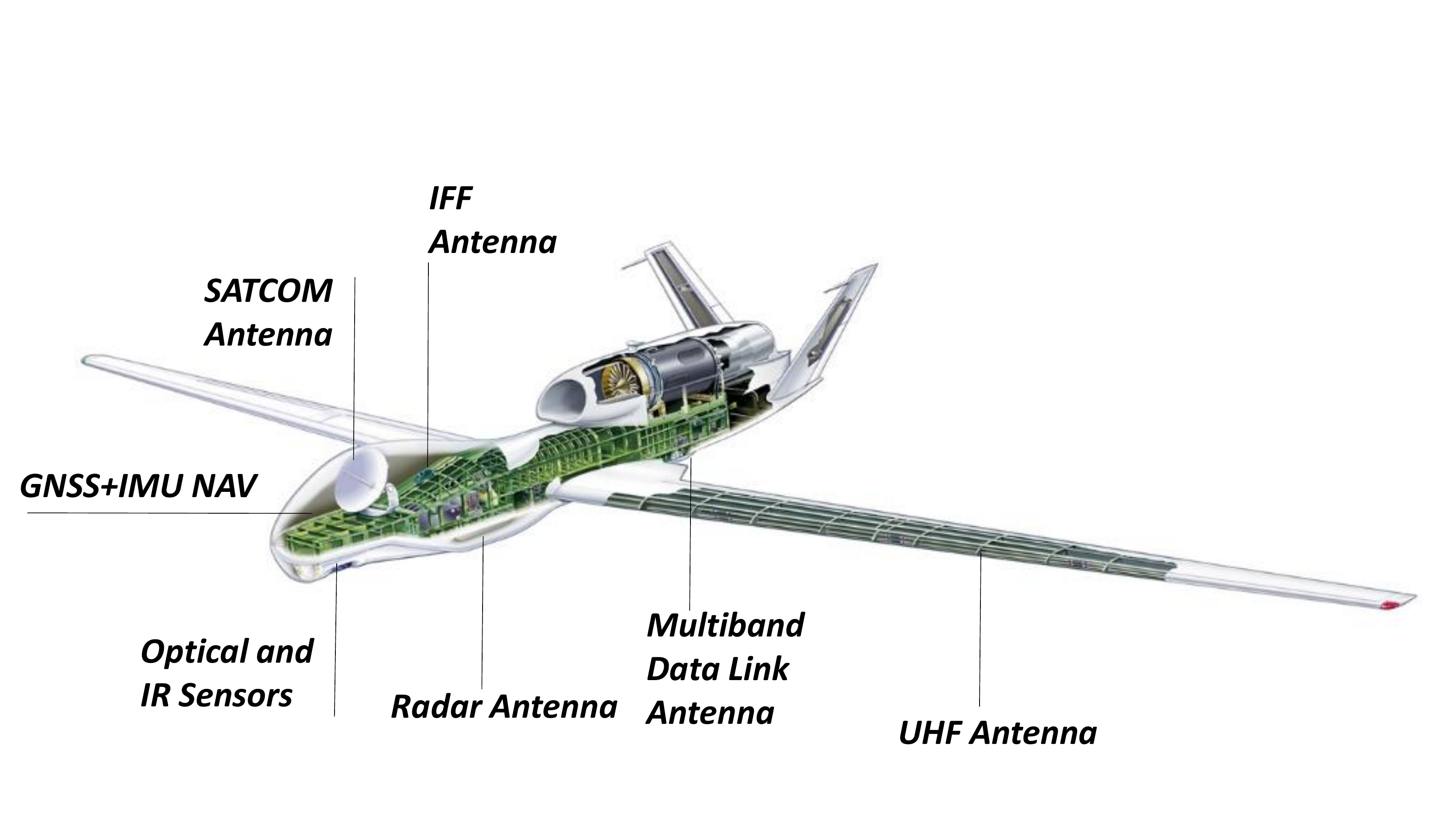}
	\caption{Sensing and Communication Components of a UAV}
	\label{fig:anatomy}
\end{figure}

UAVs are cyber-physical systems, meaning that their operations are reliant on the interaction between physical and computational elements of the system. Consequently, security of a UAV is dependent not only on the computation and communications elements and protocols, but also on the physical components of the system~\cite{banerjee2012ensuring}. This heavy entanglement of traditionally independent components requires a thorough framework for analysis of security issues in UAVs to be inclusive of the entire airframe. One obstacle in developing such a framework is the variety of UAV architectures and capabilities which makes the design of a generic model difficult. Yet, the similarity of fundamental requirements of such systems allows for generation of a high level system model for conventional types of UAVs. Figure \ref{fig:anatomy} depicts a breakdown of components in a conventional UAV. Most UAVs contain multiple communication antennas, including air to ground (ATG), air to air (ATA), satellite data link and navigation antennas, along with a set of sensors. The positioning and navigation of a UAV is typically consisted of a Global Navigation Satellite System (GNSS) receiver for accurate positioning, and an Inertial Measurement Unit (IMU) for relative positioning based on readings from kinetic sensors. This subsystem can be further extended to include air traffic monitors such as ADS-B and collision avoidance systems.

Inside the fuselage, one or more processors supervise the operation and navigation of the UAV, using the output of various radios and sensors for adjustment of electronic and mechanical parameters. This process is performed by adaptive control mechanisms, many of which are dependent on feedback loops.  Each of the elements mentioned in this section may become the subject of malicious exploitation, leading the UAV into undesirable states and critical malfunctions.  
\section{Overview of Potential Attacks}\label{related}
Table \ref{tab:table1} lists some of the uninvestigated attacks on UAS networks, categorized according to both network functionalities and cyber-physical factors. The table emphasizes on the criticality of the security problem, as the potential for vulnerability exists in every major component, ranging from the outer fuselage and antennas to network layers and application stack. This section provides an overview on the attacks listed in Table \ref{tab:table1}, and presents preliminary ideas on potential mitigating approaches and areas of research.

\vspace{15pt}
\begin{table}[h]
	\normalsize
	\begin{center}
		\caption{Cyber-Physical Attacks on UAS Networks}
		\label{tab:table1}
		\begin{tabular}{p{0.75in}|p{1.75in}}
			Component & Attacks\\
			\hline
			\\
			Sensors & Visual Navigation Jamming and Spoofing\\
			\hline
			\\
			Physical Layer & \emph{Adaptive radios:} deceptive attacks on spectrum sensing\\ 
			\\
			~ & \emph{Antennas:} Disruption and Deception of Direction of Arrival estimator, Beamnull-induced Jamming\\ 
			\\
			~ & \emph{Orientation:} Self-disruption by Induction of Defensive Maneuvers\\
			\hline
			\\
			Link Layer & Topology Inference, Topological Vulnerability of Formation to Adaptive Jamming, Routing attacks\\
			\hline
			\\
			Network Layer & Traffic Analysis, Disruption of Convergence \\
			\hline
			\\
			Air Traffic Control & ADS-B Spoofing, TCAS-Induced Collisions\\
			\hline
			\\
			Fault Handling & Manipulation of fault detection\\
			\hline
			
		\end{tabular}
		
	\end{center}
\end{table}

\subsection{Sensors and Navigation}
Absence of a human pilot from the airframe of UAVs puts the burden of observing the environment on the set of sensors onboard the aircraft. Whether autonomous or remotely piloted, sensors are the ``eyes and ears'' of the flight controller and provide the environmental measurements necessary for safe and successful completion of the mission. However, malicious exploitation of sensors in critical cyber-physical systems is widely neglected in vulnerability assessment of such systems. An attacker may manipulate or misuse sensory input or functions to trigger or transfer malware, misguide the processes dependent on such sensors, or simply disable them to cause denial of service attacks and trigger undesired fail-safe mechanisms \cite{subramanian2014sensory}.

For navigational measurements, GNSS and IMU units are traditionally used in tandem to provide accurate positioning of the aircraft. It is well-known that GNSS signals, such as GPS, are highly susceptible to spoofing attacks. The report in~\cite{wesson2013hacking} demonstrates that UAVs that only rely on commercial GPS receivers for positioning are vulnerable to relatively simple jamming and spoofing attacks, which may lead to crash or capture of the UAV by adversaries. Since the establishment of GPS, various countermeasures against GNSS spoofing have been proposed, ranging from exploitation of direction and polarization of the received GPS signal for attack detection to beamforming and statistical signal processing methods for elimination of spoofing signals~\cite{jafarnia2012gps}. However, the speed and spatial freedom of UAVs render many of the basic assumptions and criteria of such techniques inapplicable. In~\cite{humphreys2008assessing}, the authors propose the cross-examination of variations in IMU and GPS readings for detection of spoofing attacks from anomalies in fused measurements. While theoretically attractive, practical deployment of this technique requires highly reliable IMUs and adaptive threshold control for an efficient performance, which are economically undesirable for the small UAVs industry. Such practical limitations in accuracy and implementation leave this detection technique ineffective to advanced spoofing attacks, demonstrating the insufficiency of current civilian GNSS technology for mission-critical applications.

Fusion of IMU and GNSS systems with other sensors, such as video camera, may lessen the possibility of spoofing. Yet, vision-based navigation is also subject to attacks, the simplest of which is blinding the camera by saturating its receptive sensors with high intensity laser beams. A more sophisticated attack may aim for deception of the visual navigation system: In smaller areas, homogenizing or periodically modifying the texture of the terrain beneath a camera-equipped UAV may cause miscalculations of movement and orientation. Investigating the effect of such attacks on the control loop of a fused positioning system may determine the feasibility of such attacks and potential mitigation techniques.

Detection of attacks on the navigation subsystem is the basis of reactive countermeasures, such as triggering of hovering or return-to-base mechanisms. However, as the following section demonstrates, fault-handling mechanisms are also potential subjects to malicious manipulation. Robustness of the sensory and navigational subsystem against spoofing attacks may be further improved by implementation of proactive mechanisms through elimination of spoofing signals, applicability of which to UAVs is yet to be investigated.

\subsection{Fault Handling Mechanisms}
Even with the stringent reliability requirements of UAVs, mechanical and electronic subsystems of UAVs remain prone to faults due to physical damage and unpredicted state transitions. Therefore, critical UAV systems must consider the possibility of faults and implement Fault Handling mechanisms to reduce the impact of such events on the system. Typical examples of fault handling mechanisms are entering a hovering pattern when temporary faults occur,  return-to-base for persistent faults and self-destruction in the event of fatal faults such as capture or crash. In remotely operated systems, fault handling mechanisms may be triggered automatically once a certain fault is detected. This process adds yet another attack surface to UAS networks, as the fault detection mechanisms may be subject to manipulation~\cite{hartmann2013vulnerability}. For instance, if a temporary disruption of communications triggers the hovering pattern of a UAV, an adversary can jam the link to bind the motion of the aircraft, thus simplify its kinetic destruction or physical capture. A more severe case is when sensory manipulation allows the induction of capture conditions on a tactical UAV, thereby triggering its auto-destruction mechanism. 

\subsection{Air Traffic Control (ATC) and Collision Avoidance}
Integration of unmanned vehicles with national and international airspaces requires guarantees on safety and reliability of UAV operations. One major consideration in the safety of all airborne operations -manned and unmanned- is situation awareness and collision avoidance. Modern manned aircraft in the major civilian airspaces are equipped with secondary surveillance technologies such as Automatic Dependent Surveillance - Broadcast (ADS-B), which allow each aircraft to monitor the air traffic in their vicinity. This information, along with other available means of traffic monitoring, provide situation awareness to the Traffic advisory and Collision Avoidance System (TCAS), which monitors the risk of collision with other aircraft and generates advisories on how to prevent collisions.

With the growing interest in large-scale deployment of UAVs, implementation of similar technologies in UAS is crucial. Recent literature contain several proposals on TCAS and ATC solutions for UAVs, many of which are based on adaptation of ADS-B and commercial TCAS protocols. From a security point of view, this approach suffers from several critical vulnerabilities, rendering it unfeasible for mission-critical UAS applications. Firstly, ADS-B is an insecure protocol by design~\cite{kim2012cyber}. Lack of authentication and the unencrypted broadcast nature of this protocol make room for relatively simple attacks, ranging from eavesdropping to manipulation of air traffic data by jamming or injection of false data. Consequently, a TCAS system relying on ADS-B can produce erroneous results and advisories, leading to unwanted changes in the flight path or in the worst scenario, collisions.

\begin{figure}
	\centering
	\includegraphics[width=\linewidth]{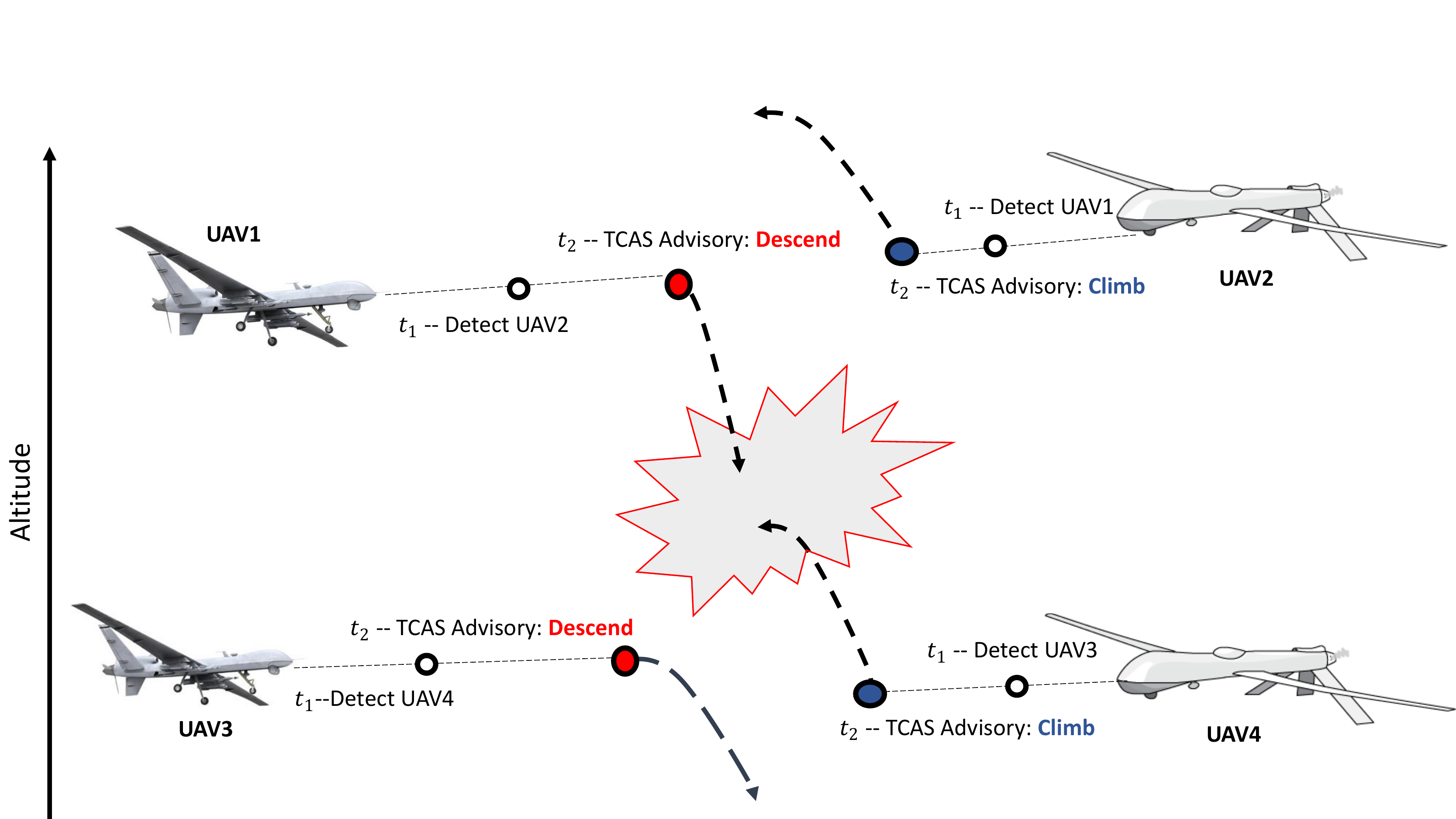}
	\caption{Example of TCAS-Induced Collision in a 4 Airplane Scenario}
	\label{fig:tcas}
\end{figure}

Also, TCAS is shown to be susceptible to a flaw known as ``TCAS-Induced Collisions''~\cite{tang2015causal}. Common implementations of TCAS are not equipped with prediction capabilities to foresee the longer-term effect of an advisory that they produce. In dense traffic conditions, certain scenarios may cause the TCAS to generate advisories that lead to a state where avoidance of collision is not possible. Hence, an adversary capable of manipulating the traffic data can intentionally orchestrate conditions leading to TCAS-induced collisions. Authors of \cite{tang2015causal} provide an example of this flaw for a 4 airplane scenario, as illustrated in Figure \ref{fig:tcas}. In this scenario, UAV1 and UAV2 are initially in a collision path, hence the TCAS in each generates a collision avoidance advisory to descend and climb, respectively. At a lower altitude, the same situation holds for UAV3 and UAV4, causing UAV4 to climb, which puts UAV1 and UAV4 on a collision path. Even though that TCAS does not fail to generate new correction advisories in both UAVs, but the advisory is no longer practical as there is not enough time before the collision to implement the new path.

\subsection{Physical Layer}
Typical UAVs require multiple radio interfaces to retain continuous connectivity with essential links to satellite relays, ground control stations and other UAVs. This degree of complexity, along with the physical and mechanical characteristics of UAVs, widen the scope of potential vulnerabilities and enable multiple attacks that are specific to UAS networks. This section presents a discussion on some of such attacks on the physical layer of UAV nodes.

\textbf{\textit{1. Adaptive Radios:}}
As the operational environment of UAS networks is highly dynamic, sustained and reliable communications necessitates the employment of radios that are capable of adjusting to changes in propagation and links conditions. Depending on the operational requirements, this adaptability may apply to any of the physical layer parameters such as transmit power, frequency, modulation, and configuration of antennas. The procedure responsible for controlling these parameters must essentially rely on environmental inputs, which can be manipulated by adversaries to result in undesirable configurations. This issue is analogous to deceptive attacks on the spectrum sensing process of cognitive radio networks, for which various mitigation techniques have been proposed based on anomaly detection and fusion of distributed measurements~\cite{bhattacharjee2013vulnerabilities}. However, the rapid variation of conditions in a UAS network may lead to situations where determination of a baseline for anomaly detection is not practical. The same consideration also develops a necessity for rapid adjustments, which limits the acceptable amounts of redundancy and overhead. Similarly, deployment of airborne nodes in hostile environments further reduces the feasibility of relying on real-time collaboration between distributed sensors. Therefore, such countermeasures will not be sufficient for agile UAS radios and novel solutions must be tailored according to the unique requirements of airborne networks.

\textbf{\textit{2. Antennas:}}
The current trend in antenna selection for UAV radios is favored towards omnidirectional antennas, defined by their relatively homogeneous reception and transmission in all directions of the horizontal or vertical planes. This feature simplifies communications in mobile nodes, as the homogeneity of gain eliminates the need for considering the direction of transmissions. On the other hand, the indiscriminate nature of omnidirectional antennas extends the attack surface for eavesdroppers and jammers, since they also do not need to tune towards the exact direction of radios to implement their attacks. A countermeasure against this class of attack is the utilization of directional antennas, which can only communicate in certain directions and are ``blind'' to others. Besides their higher security, other advantages of directional antennas include longer transmission ranges and spatial reuse, thus providing a higher network capacity. One downside associated with this approach is the inevitable escalation of overhead. Maintaining directional communications in highly mobile networks is a complex and costly task, as it requires real-time knowledge of other nodes' positions, as well as employment of antennas capable of reconfiguring their beam patterns.

To overcome the disadvantages of these two approaches, a midway solution combining the simplicity of omnidirectional radios and spatial selectivity of directional antennas can be actualized in the form of beamforming antenna arrays. Such antennas are capable of detecting the Direction of Arrival (DoA) of individual signals. This measurement, along with other system parameters, are then used to electronically reconfigure the radiation pattern and directionality of the antenna array. Beamforming has been studied as a mitigation technique against jamming attacks, as it allows spatial filtering of the jammer's signals by adjusting the antenna pattern such that a null is placed towards the direction of the jammer~\cite{bhuniaperformance}. The accuracy and efficiency of this technique depends on correct detection of the jamming signal, as well as the resolution of beamformer's DoA estimations. An adversary may attack the DoA estimator by shaping its jamming signals to mimic waveforms of a nearby legitimate node, thus avoiding detection or causing false detections.

Another attack scenario exploits the process of beamnulling itself. In an ad hoc UAS network, beamnulling must be implemented in a distributed fashion to allow targeted nodes to retain or regain connectivity with the network independently. Due to lack of coordination, nulls created by one node towards a jammer may also null the direction of legitimate signals. Depending on the mobility model and formation of the network, an adversary may deploy multiple mobile jammers with strategically controlled trajectories to manipulate the DoA measurements, and eventually cause the network to null more of its legitimate links than is necessary. In certain conditions, the adversary can maximize the efficiency of jamming attacks by persistently manipulating the distributed beamnulling mechanism in such a way that its solution converges towards a maximally disconnected state. Analytical studies into feasibility criteria of this attack may produce insights into possible countermeasures and mitigation techniques. 

\textbf{\textit{3. Orientation:}}
As depicted in figure \ref{fig:anatomy}, a conventional UAV employs multiple fixed antennas on different sides, each of which is dedicated to a certain application. Consider the ATG antenna which is placed on the lower side of the UAV. As discussed previously, if the UAV performs a half-roll maneuver or ascends with a steep climb angle, the ATG antenna is no longer capable of communicating with the ground antenna and therefore the ATG link is lost. This issue can be exploited for jamming in UAS Networks that employ the spatial retreat as a mitigation technique. By observing the reaction of the nodes to jamming attacks, an adversary may infer their reformation strategy, and adapt its attack such that the defensive reformation of certain nodes leads to the loss of some links due to the new orientation of antennas. 

\subsection{Link Layer and Formation}
Similar to generic multihop wireless networks, the topology of a UAS network is determined based on the location of UAVs relative to each other: UAVs closer than a threshold can directly communicate with each other, while those that are farther must utilize relay nodes to reach their destination. Knowledge of the topology of a network allows adversaries to optimize attacks by analyzing the structure of their target and determine the most vulnerable regions by identifying nodes whose disconnection incur the maximum loss of connectivity in the network. Even though the effect of topology on the resilience of the network is widely studied, the proposed mitigation techniques fail to provide practical solutions for UAS networks. A class of such solutions are based on a “security by obscurity” approach, suggesting the employment of covert communications between nodes to hide the topology of the network from adversaries. Besides the undesirable overhead of this approach in terms of decreased network throughput and increased processing costs, it has been shown that the topology of such networks can be estimated with a high degree of accuracy via timing analysis attacks~\cite{Vahid}.
Therefore, hiding the topology may not serve as a reliable solution in mission critical scenarios.

An alternative mitigation technique is adaptive control of the topology~\cite{zhu2011attack}. In this approach, detection of a jamming attack triggers a reformation process during which the nodes of a UAS network change their positions to retain connectivity. A fundamental assumption of this approach is the ability of the nodes to detect and localize attacks, which may not always be practical. A promising area of further investigation is the problem of minimizing the topological vulnerability to targeted jamming attacks. Development of real-time and distributed formation control techniques that consider this optimization problem may lead to highly efficient techniques for ensuring dynamic resilience of mission-critical UAS networks.

A mitigation technique against topology inference attacks is randomization of transmission delays. It is expected that introducing randomness in forwarding delays weakens the observed correlation between connected hops, and therefore reduces the accuracy of timing analysis attacks. However, the high mobility of UAS networks and the consequent requirement for minimal latency limit the maximum amount of delay permissible in such networks. This constraint limits the randomness of the forwarding delays, which may neutralize the effect of mitigation technique. A potential alternative for delay randomization is transmission of decoy signals to perturb the adversary's correlation analysis. This proposal may be extended by incorporating it in topology control, such that the resultant formation is optimized for decoy transmissions in a way that spatial distribution of traffic in the network appears homogeneous to an outside observer, thereby inducing an artificial correlation between all nodes in the network. To the extent of authors' knowledge, the feasibility, overhead and optimal implementation of this approach are yet to be analytically and experimentally studied.

\subsection{Network Layer}
The impact of high mobility in UAS networks is greatly accentuated in the network layer. Speed and frequency of changes in the topology of a UAS network give rise to many challenges that are still active subjects of research. Yet, studies on security of routing mechanisms tend to follow the tradition of equating UAS networks with MANETs. Indeed, the unique features of unmanned airborne networks generate a set of challenges in the network layer that do not match the criteria of conventional MANETs. The highly dynamic nature of UAS networks, as well as stringent requirements on latency, necessitate novel routing mechanisms capable of calculating paths in rapidly changing topologies. A survey of the state of the art in this area is presented in~\cite{bekmezci2013flying}. The proposed methods may be prone to potential vulnerabilities, and the demand for a detailed technical analysis and comparison of these proposals in terms of their security is yet to be fulfilled.

Similar to the link layer, the routing layer of UAS networks is also vulnerable to traffic analysis attacks, aiming to infer individual flows, as well as source-destination pairs of end-to-end connections. Various mitigation techniques against such attacks have been proposed~\cite{kong2007identity}, many of which rely on traditional approaches such as mixing and decoy transmissions. As such techniques require addition of redundancies and overhead to the UAS networks, a comprehensive feasibility analysis and optimal design of the corresponding defense strategies is vital, but not yet available to the research community. 

Mobile routing in UAS networks is a surface for attacks on convergence of the network. As discussed, the topology of unmanned airborne networks is subject to manipulation by adversarial actions such as exploitation of adaptive formation control and jamming attacks. Also, many of the recently proposed routing mechanisms for airborne networks rely on global knowledge of the geographical positions of every node in the network, which may also be prone to manipulation. A sophisticated adversary may be able to design a strategic combination of topological perturbation and sensor manipulations to prevent or slow the convergence of routing in the network. Investigation of this attack in terms of feasibility, as well as potential countermeasures may prove to be valuable for efficient protection of UAS networks operating in hostile environments. 
\section{Conclusions}\label{conclusion}
The cyber-physical nature of UAVs demand an extension to the scope of ordinary vulnerability analysis for such systems. In addition to threats in the electronic and computational components, a largely overlooked class of vulnerabilities is fostered by the interactions between the mechanical elements and the computational subsystems. Pondering on the list of critical attacks presented in this paper, an alarming conclusion can be drawn: serious threats still remain unmitigated not only in every networking component of UAS communications, but also in the interdependency of the network and other components, including sensors and physical elements of UAVs. Considering the seriousness of open issues in the cyber-physical aspects of UAVs, a successful move towards the age of mainstream unmanned aviation cannot be envisioned without remedying the void of effective solutions for such critical challenges.

\vspace{0pt}

\begin{thebibliography}{10}
	
	\bibitem{kim2012cyber}
	A.~Kim, B.~Wampler, J.~Goppert, I.~Hwang, and H.~Aldridge, ``Cyber attack
	vulnerabilities analysis for unmanned aerial vehicles,'' {\em Infotech@
		Aerospace}, 2012.
	
	\bibitem{javaid2012cyber}
	A.~Y. Javaid, W.~Sun, V.~K. Devabhaktuni, and M.~Alam, ``Cyber security threat
	analysis and modeling of an unmanned aerial vehicle system,'' in {\em
		Homeland Security (HST), 2012 IEEE Conference on Technologies for},
	pp.~585--590, IEEE, 2012.
	
	\bibitem{banerjee2012ensuring}
	A.~Banerjee, K.~K. Venkatasubramanian, T.~Mukherjee, and S.~K.~S. Gupta,
	``Ensuring safety, security, and sustainability of mission-critical
	cyber--physical systems,'' {\em Proceedings of the IEEE}, vol.~100, no.~1,
	pp.~283--299, 2012.
	
	\bibitem{subramanian2014sensory}
	V.~Subramanian, R.~Beyah, {\em et~al.}, ``Sensory channel threats to cyber
	physical systems: A wake-up call,'' in {\em Communications and Network
		Security (CNS), 2014 IEEE Conference on}, pp.~301--309, IEEE, 2014.
	
	\bibitem{wesson2013hacking}
	K.~Wesson and T.~Humphreys, ``Hacking drones,'' {\em Scientific American},
	vol.~309, no.~5, pp.~54--59, 2013.
	
	\bibitem{jafarnia2012gps}
	A.~Jafarnia-Jahromi, A.~Broumandan, J.~Nielsen, and G.~Lachapelle, ``Gps
	vulnerability to spoofing threats and a review of antispoofing techniques,''
	{\em International Journal of Navigation and Observation}, vol.~2012, 2012.
	
	\bibitem{humphreys2008assessing}
	T.~E. Humphreys, B.~M. Ledvina, M.~L. Psiaki, B.~W. O’Hanlon, and P.~M.
	Kintner~Jr, ``Assessing the spoofing threat: Development of a portable gps
	civilian spoofer,'' in {\em Proceedings of the ION GNSS international
		technical meeting of the satellite division}, vol.~55, p.~56, 2008.
	
	\bibitem{hartmann2013vulnerability}
	K.~Hartmann and C.~Steup, ``The vulnerability of uavs to cyber attacks-an
	approach to the risk assessment,'' in {\em Cyber Conflict (CyCon), 2013 5th
		International Conference on}, pp.~1--23, IEEE, 2013.
	
	\bibitem{tang2015causal}
	J.~Tang, {\em Causal models for analysis of TCAS-induced collisions}.
	\newblock PhD thesis, Universitat Aut{\`o}noma de Barcelona, 2015.
	
	\bibitem{bhattacharjee2013vulnerabilities}
	S.~Bhattacharjee, S.~Sengupta, and M.~Chatterjee, ``Vulnerabilities in
	cognitive radio networks: A survey,'' {\em Computer Communications}, vol.~36,
	no.~13, pp.~1387--1398, 2013.
	
	\bibitem{bhuniaperformance}
	S.~Bhunia, V.~Behzadan, P.~A. Regis, and S.~Sengupta, ``Performance of adaptive
	beam nulling in multihop ad hoc networks under jamming,'' {\em IEEE
		International Symposium on Cyberspace Safety and Security (CSS 2015), New
		York}.
	
	\bibitem{Vahid}
	V.~Behzadan and S.~Sengupta, ``Real-time inference of topological structure and
	vulnerabilities for adaptive jamming against tactical ad hoc networks,'' {\em
		Under Review in Elsevier Journal of Computer and System Sciences}, 2016.
	
	\bibitem{zhu2011attack}
	M.~Zhu and S.~Mart{\'\i}nez, ``Attack-resilient distributed formation control
	via online adaptation,'' in {\em Decision and Control and European Control
		Conference (CDC-ECC), 2011 50th IEEE Conference on}, pp.~6624--6629, IEEE,
	2011.
	
	\bibitem{bekmezci2013flying}
	I.~Bekmezci, O.~K. Sahingoz, and {\c{S}}.~Temel, ``Flying ad-hoc networks
	(fanets): A survey,'' {\em Ad Hoc Networks}, vol.~11, no.~3, pp.~1254--1270,
	2013.
	
	\bibitem{kong2007identity}
	J.~Kong, X.~Hong, and M.~Gerla, ``An identity-free and on-demand routing scheme
	against anonymity threats in mobile ad hoc networks,'' {\em Mobile Computing,
		IEEE Transactions on}, vol.~6, no.~8, pp.~888--902, 2007.
	
\end{thebibliography}

\end{document}